\documentclass[%
reprint,
showpacs,
showkeys,
nofootinbib,
nobibnotes,
amsmath,fixed
amssymb,
aps, 
superscriptaddress
]{revtex4-1}

\usepackage[english]{babel}
\usepackage[utf8x]{inputenc}
\usepackage[T1]{fontenc}
\usepackage{lineno}
\usepackage{amsmath}
\usepackage{graphicx}
\usepackage[export]{adjustbox}

\usepackage[colorlinks=true,linkcolor=blue,citecolor=blue,urlcolor=blue]{hyperref}

\usepackage{epstopdf}
\usepackage{amssymb}
\usepackage{color}
\usepackage{url}
\usepackage[justification=justified]{subcaption}

\usepackage{float}
\restylefloat*{figure}
\usepackage{placeins}

\usepackage{dcolumn}
\usepackage{bm} 

\usepackage{siunitx}
\usepackage{textcomp} 
\usepackage{multirow}
\usepackage{tabularx} 
\usepackage{ragged2e}
\usepackage[overload]{textcase}

\begin{document}


\title{A search for new physics in low-energy electron recoils from the first LZ exposure}

\author{J.~Aalbers}
\affiliation{SLAC National Accelerator Laboratory, Menlo Park, CA 94025-7015, USA}
\affiliation{Kavli Institute for Particle Astrophysics and Cosmology, Stanford University, Stanford, CA  94305-4085 USA}

\author{D.S.~Akerib}
\affiliation{SLAC National Accelerator Laboratory, Menlo Park, CA 94025-7015, USA}
\affiliation{Kavli Institute for Particle Astrophysics and Cosmology, Stanford University, Stanford, CA  94305-4085 USA}

\author{A.K.~Al Musalhi}
\affiliation{University of Oxford, Department of Physics, Oxford OX1 3RH, UK}

\author{F.~Alder}
\affiliation{University College London (UCL), Department of Physics and Astronomy, London WC1E 6BT, UK}

\author{C.S.~Amarasinghe}
\affiliation{University of Michigan, Randall Laboratory of Physics, Ann Arbor, MI 48109-1040, USA}

\author{A.~Ames}
\affiliation{SLAC National Accelerator Laboratory, Menlo Park, CA 94025-7015, USA}
\affiliation{Kavli Institute for Particle Astrophysics and Cosmology, Stanford University, Stanford, CA  94305-4085 USA}

\author{T.J.~Anderson}
\affiliation{SLAC National Accelerator Laboratory, Menlo Park, CA 94025-7015, USA}
\affiliation{Kavli Institute for Particle Astrophysics and Cosmology, Stanford University, Stanford, CA  94305-4085 USA}

\author{N.~Angelides}
\affiliation{Imperial College London, Physics Department, Blackett Laboratory, London SW7 2AZ, UK}

\author{H.M.~Ara\'{u}jo}
\affiliation{Imperial College London, Physics Department, Blackett Laboratory, London SW7 2AZ, UK}

\author{J.E.~Armstrong}
\affiliation{University of Maryland, Department of Physics, College Park, MD 20742-4111, USA}

% ...

\collaboration{LZ Collaboration}

\begin{abstract}
\noindent
The LUX-ZEPLIN (LZ) experiment is a dark matter detector centered on a dual-phase xenon time projection chamber. We report searches for new physics appearing through few-keV-scale electron recoils, using the experiment's first exposure of 60 live days and a fiducial mass of 5.5~t.  The data are found to be consistent with a background-only hypothesis, and limits are set on models for new physics including solar axion electron coupling, solar neutrino magnetic moment and millicharge, and electron couplings to galactic axion-like particles and hidden photons.  Similar limits are set on weakly interacting massive particle (WIMP) dark matter producing signals through ionized atomic states from the Migdal effect.
\end{abstract}


\maketitle
\enlargethispage{\baselineskip} 

\section{Introduction}
\label{sec:intro}

Liquid xenon (LXe) time projection chambers~(TPCs) are the most sensitive technology searching for weakly interacting massive particle~(WIMP) dark matter via keV-scale nuclear recoils (NRs)~\cite{LUX_2016DMresults,XENON_SI-DM_2018,PandaX-4T_SI-DM_2021}. Detectors of this type are also sensitive to numerous beyond the standard model (BSM) physics processes that would be expected to generate signals in the electron recoil (ER) channel~\cite{XENON_ERlimits_2022,LZ_LowERsensitivity}. Recently, the LUX-ZEPLIN (LZ) collaboration presented WIMP search results for its first science exposure, using a fiducial mass of 5.5~tonnes over a period of 60~live days~\cite{LZ_WIMPresults}. Here, we present searches for new physics signatures in the ER channel using the same exposure, and employing both the same data selection criteria and the same models of detector response.  We search for evidence of recoiling electrons from interactions with solar axions, axion-like particles, and hidden photons, as well as non-standard interactions with solar neutrinos coming from magnetic moments and millicharge. Additionally, we search for WIMP-xenon scattering via the Migdal effect, which produces electrons from atomic ionization. These searches exploit sensitivities to energy depositions between $\sim$1 and $\sim$15~keV, with signals expressed as either a line-feature or characteristic spectral shape within this range. The method of statistical inference uses the same methods as the WIMP analysis~\cite{LZ_WIMPresults}, with one change: time is added as an additional parameter of interest to take advantage of well-understood temporal variation in ER backgrounds, specifically the decay of $^{37}$Ar and $^{127}$Xe. In all tested new physics models, the data agree well with the background-only hypothesis, allowing exclusion of the specific signal model at a defined level of confidence. 

\begin{figure}[htbp]
    \centering
    \includegraphics[width=0.96\linewidth]{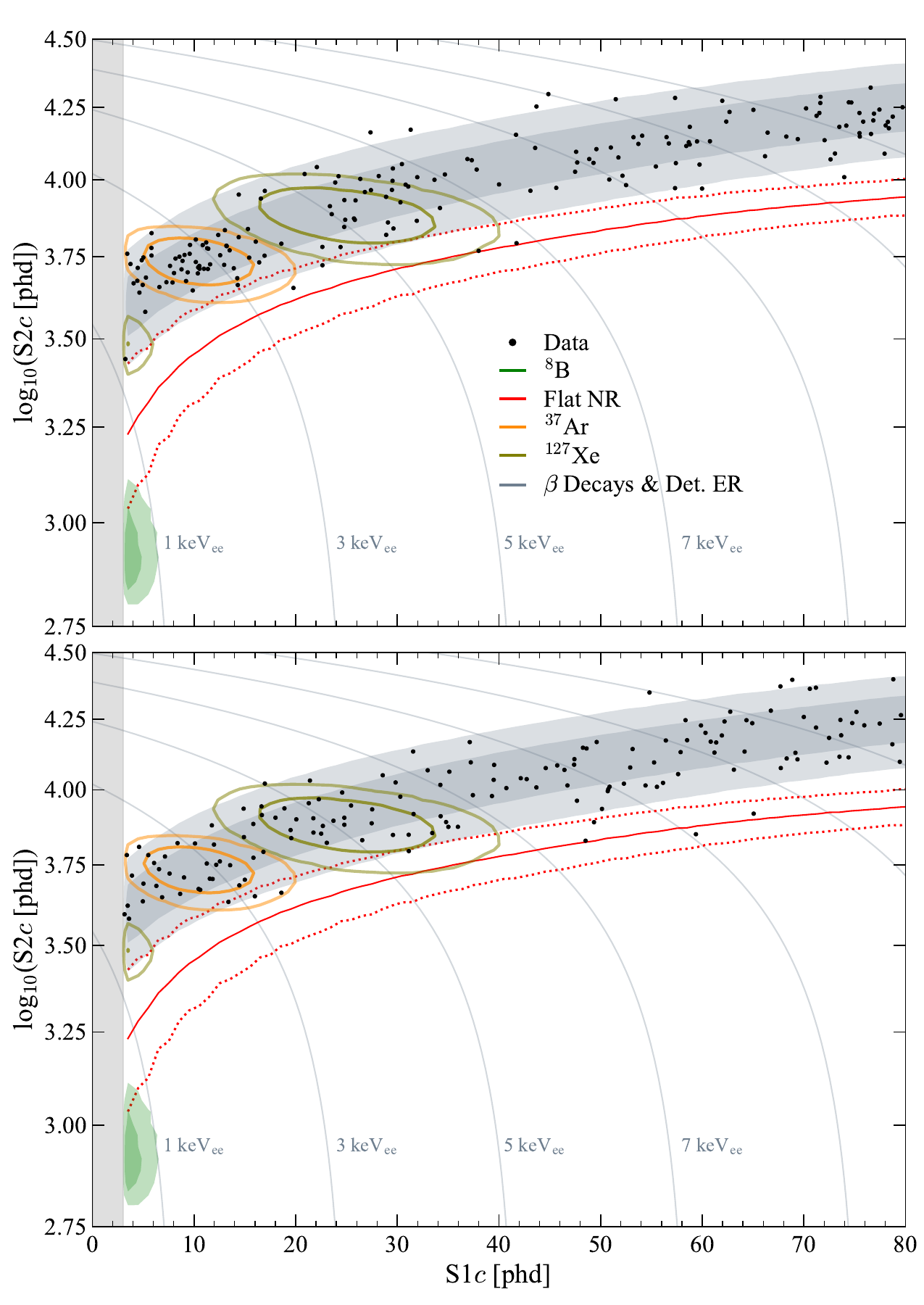}
    \caption{The SR1 search data (black points) is plotted in the \{S1$c$, log$_{10}$S2$c$\} space, after all cuts and selections are applied.  For illustration purposes, the exposure is shown separated into two periods of equal livetime (top panel is the first half of SR1, bottom panel is the second half). In both panels, the 1$\sigma$ and 2$\sigma$ regions are indicated for various background model components:  $^{37}$Ar (orange contours), $^{127}$Xe (green contours), $^{8}$B (filled green), and the broad-spectrum ER background encompassing $^{212}$Pb, $^{214}$Pb, $^{85}$Kr, and external gammas (filled gray). The solid red line indicates the median, the dashed the 10\% and 90\%, quantiles of a flat NR background. Thin gray lines indicate contours of constant ER energy, with a spacing of 2 keV$_{\mathrm{ee}}$.  A reduction in $^{37}$Ar rate is the dominant change between the two time periods.}
    \label{fig:s1logs2data}
\end{figure}

\begin{figure}[htbp]
  \centering
  \includegraphics[width=0.96\linewidth]{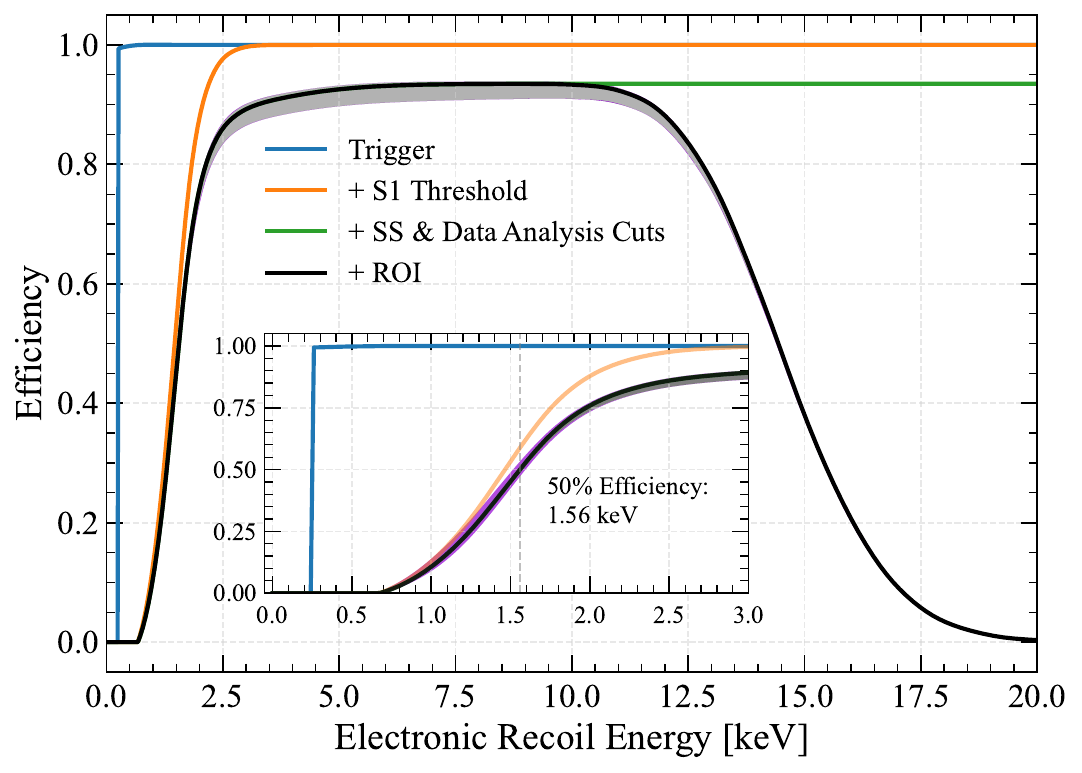}
  \caption{Signal efficiency as a function of simulated true ER energy for the trigger (blue), the $\ge$3-fold coincidence and $>$3~phd threshold on S1$c$ (orange), single-scatter (SS) reconstruction and analysis cuts (green), and the search ROI in S1 and S2 (black). The inset shows the low energy behavior, with the dotted line at 1.56~keV marking 50\% efficiency. The data quality acceptance error band (gray) is assessed using AmLi and tritium data. The \textsc{nest} model uncertainties (purple) are discussed in Appendix A. }
  \label{fig:efficiency}
\end{figure}

\section{Detector, data and selection treatment}
\label{sec:detector}

The LZ detector is described in Refs.~\cite{LZ_TDR, LZ_detector, LZ_WIMPresults}. It is located within the Davis Cavern at the Sanford Underground Research Facility (SURF) in Lead, South Dakota. A 1100~m rock overburden provides shielding from cosmic ray muons equivalent to 4300~m of water, reducing their flux by a factor of $3\times10^6$~\cite{Mei_2010,Kudryavtsev_2009}. 
The central region of the LXe TPC is shielded from ambient radiation of the local environment by surrounding layers of active and passive materials:  an instrumented LXe ``skin'' detector, an instrumented outer detector containing 17~t of gadolinium-loaded liquid scintillator, and a surrounding tank filled with 238~t of ultra-pure water.  Radiopurity requirements were maintained through construction~\cite{LZ_cleanliness}, and the observed background rates have been described in Ref.~\cite{LZ_backgrounds}.

The central TPC consists of a vertical cylinder 145.6~cm in diameter and height (when cold), lined with reflective PTFE.  Titanium rings and four wire-mesh electrode grids form a field cage to establish a near-uniform 193~V/cm vertical electric field throughout the drift region. The TPC is instrumented with two arrays of Hamamatsu R11410-20-type 3-inch photomultiplier tubes (PMTs), 253 PMTs at the top and 241 PMTs at the base.  

Energy depositions occurring within the TPC produce vacuum ultraviolet (VUV) scintillation photons (S1) and ionization electrons which drift under the electric field to the liquid surface. Gate and anode electrodes establish an extraction field of 7.3~kV/cm in gas just above at the center of the surface, drawing the ionization electrons into the gas where secondary scintillation is produced (S2). The ratio of S2 and S1 signal amplitudes may be used to discriminate between NR and ER type energy depositions.   

Using mono-energetic calibration sources dispersed into the LXe ($^{83m}$Kr and $^{131m}$Xe), the position dependence of the S1 signal can be observed and accounted for by normalizing all positions to the geometric center of the detector; this `corrected' value of S1 is called S1$c$. Similarly, the S2 signal is normalized to the radial center and top (shortest drift time) of the TPC; this corrected value is called S2$c$. As reported in Ref.~\cite{LZ_WIMPresults}, the size of the S1 corrections is on average 9\%, and the size of the S2 corrections is on average 11\% in the \{x, y\} plane. The S2 correction in z (due to time-dependent variation in the probability for electrons to be captured by electronegative impurities) averages 7\%. Corrected parameters are uniform across the TPC to within 3\%.

After the search period concluded, tritium in the form of $^3$H-labeled CH$_4$ was used as a calibration source.  Detector and Xe response parameters of \textsc{nest} 2.3.7~\cite{NEST_237} are tuned to match the median and widths of $^3$H calibration data in $\{$S1$c$, log$_{10}$S2$c$$\}$ space, and simultaneously match the reconstructed energies of the $^{83m}$Kr (41.5 keV), $^{129m}$Xe (236 keV), and $^{131m}$Xe (164 keV) peaks. Using the S1$c$ and S2$c$ quantities, the photon detection efficiency, $g_1$, is found to be 0.114$\pm$0.002~phd/photon and the gain of the ionization channel, $g_2$, to be 47.1$\pm$1.1~phd/electron. Here, phd refers to photons detected, the signal size accounting for the double photoelectron effect in response to VUV photons~\cite{Faham, Paredes2015}. 

Data reported here are from Science Run 1, which ran from 23 Dec 2021 to 11 May 2022 under stable detector conditions.  Event selection follows Ref.~\cite{LZ_WIMPresults}, resulting in a total livetime of 60$\pm$1~d and a fiducial volume LXe mass of 5.5$\pm$0.2~t. A region of interest (ROI), within which the ER response is well-constrained by a tritium calibration, is defined for S1$c$ in the range 3--80~phd, uncorrected S2 greater than 600~phd ($\gtrsim$10 extracted electrons), and S2$c$ less than 10$^5$~phd.  The data of Science Run 1 is shown in the \{S1$c$, log$_{10}$S2$c$\} plane in Fig.~\ref{fig:s1logs2data}, along with contours indicating the background components considered in this analysis. The efficiency as a function of true ER energy, shown in Fig.~\ref{fig:efficiency}, is assessed on a simulated ER dataset.  Data selection efficiency is measured using calibration data including tritium and AmLi.

For this analysis, the detector tuning, data selections, and data quality acceptances were defined or calculated using a combination of dedicated calibration datasets, simulations, and/or side-band analyses so as to mitigate bias in data selection and modeling. The primary side-band dataset was formed by selecting data at energies outside the ROI, {\em e.g.} S1$c$~\textgreater~80~phd. Constraints on background model rates, as discussed in detail in Ref.~\cite{LZ_backgrounds}, were calculated using measurements of populations in high energy side-bands and/or with dedicated simulations. More detailed descriptions of analysis cuts and methods are given in Ref.~\cite{LZ_WIMPresults}.

\section{Signal Models}
\label{sec:signalModels}

In this section we discuss several BSM physics scenarios that would produce low energy ER signals in the active LXe volume. Each signal model is characterized by its energy spectrum and integrated interaction rate as a function of the relevant cross-section, coupling constant, or other physical parameter. A brief description of the theoretical motivation and production mechanism for each signal is given. 

To translate recoil spectra (in true energy) to corresponding \{S1$c$, log$_{10}$S2$c$\} observables, energy spectra for each physics signal are passed as inputs through the simulation chain utilizing \textsc{nest} and the detector response model, as tuned to calibrations. As with the modelling of backgrounds, the same suite of data selections and data quality acceptances has been applied to these signal models in the reconstructed observable space. These simulated signal distributions in \{S1$c$, log$_{10}$S2$c$\} space are then employed as probability distribution functions (PDFs) for statistical inference.  

The S1$c$ and S2$c$ amplitudes can be combined into a metric of reconstructed energy given by E$_{\mathrm{rec}} = \mathrm{W}($S1$c/g_1 + $S2$c/g_2)$, where the signals are assumed to be of ER origin and where the work function W=13.5~keV/quantum is assumed \cite{NEST_2021}.  This simple reconstructed energy quantity is used for the contours of Fig.~\ref{fig:s1logs2data} and the solid spectra of Fig.~\ref{fig:EnergySpectra}.  This reconstructed energy metric can be improved in both resolution and accuracy by first deweighting the S1$c$ contribution to the sum, and then by rescaling the overall sum to correct its mean.  This optimized E$_{\mathrm{rec}}$ quantity is described in Ref.~\cite{NEST_2021} (Section 3.1.2), demonstrated on experimental data in Ref.~\cite{Ar37pixey}, and illustrated for one example signal model in Fig.~\ref{fig:EnergySpectra}.  While an optimized E$_{\mathrm{rec}}$ quantity would provide greater sensitivity to new physics searches, any 1D energy metric suffers some loss in sensitivity compared to the 2D \{S1$c$, log$_{10}$S2$c$\} space used in this work for statistical inference.

\subsection{Electromagnetic Properties of Solar Neutrinos}

Under the Standard Model (SM), neutrinos are electrically neutral but express tiny electromagnetic couplings via radiative corrections. The electromagnetic properties considered in this work are magnetic moment $\mu_{\nu}$ and millicharge $q_{\nu}$.  These moments can be parameterized in terms of the Bohr Magneton, $\mu_{B}$, and electron charge, $e_{0}$, respectively.  Detection of these electromagnetic moments at the scales currently testable in experiment would provide evidence of beyond the standard model (BSM) physics and distinguish between the Dirac or Majorana nature of the neutrino~\cite{Bell_2005,BELL_2006,Giunti_2015}.

Following \cite{Giunti_2015}, the electromagnetic terms add to the electroweak neutrino-electron scattering cross-section (for an unbound electron) as 
\begin{equation}
\label{eqn:NuRates}
\begin{split}
    \frac{d\sigma}{dE_r} \simeq &\left(\frac{d\sigma}{dE_r}\right)_{EW}\\ 
    &+ \frac{\pi \alpha^{2}}{m_{e}^2} \left(\frac{1}{E_r} - \frac{1}{E_{\nu}}  \right) \left( \frac{\mu_{\nu}}{\mu_{B}} \right)^2 \\
    &+ \frac{2\pi\alpha}{m_{e}} \left( \frac{1}{E_{r}^2}\right)q_{\nu}^2,
\end{split}
\end{equation}
where $\alpha$ is the fine structure constant, $m_{e}$ the electron mass, $E_r$ the energy of the recoiling electron, and $E_{\nu}$ is the neutrino energy. Given the few-keV scale of $E_r$ in this work, $E_r$ $\ll$ $E_{\nu}$, meaning that the magnetic moment and  millicharge differential rates fall as $E_r^{-1}$ and $E_r^{-2}$ respectively, as seen in Fig.~\ref{fig:EnergySpectra}. In both cases the recoil rate scales with the square of their respective electromagnetic moment (either $\mu_{\nu}^{2}$ or $q_{\nu}^2$).

Corrections are required to Eq.~\ref{eqn:NuRates} for the bound state of the Xe electrons. This work relies on Ref.~\cite{Hsieh_2019} for the modeling of these effects, in which a relativistic random phase approximation (RRPA) calculation is carried out specific to electrons in Xe atomic states. These bound-state effects lead to small decreases in the rate at the lowest energies and, as in the photo-electric effect, cause discontinuities in the spectrum at the electron shell energies. 

LZ is sensitive to the `effective' electromagnetic properties of the solar neutrino flavor-mixture.  From the additive nature of the terms in Eq.~\ref{eqn:NuRates}, neutrino couplings via electromagnetic moments can be treated as separate signals, added on top of the standard electroweak process.  Sufficiently high values of either $\mu_{\nu}$ or $q_{\nu}$ lead to excess rates above background peaking at threshold, falling as $E_r^{-1}$ and $E_r^{-2}$, respectively.  Following Ref.~\cite{Hsieh_2019} and solar neutrino fluxes as described in Ref.~\cite{Vinyoles_2017}, only the three neutrino fluxes which dominate sensitivity are considered (pp, $^7$Be, CNO).

\subsection{Solar Axions}

The axion is a pseudo-scalar Nambu-Goldstone boson that results from the Peccei-Quinn mechanism~\cite{PQ_mech} introduced to interpret the small neutron electric dipole moment, often termed the strong CP problem. In the Peccei-Quinn mechanism, an additional U(1) global chiral symmetry is included in the QCD Lagrangian, representing a dynamical field; axions are the $\mu$eV/c$^{2}$-scale particle associated with this field, with couplings to leptons, hadrons, and photons. With these properties, axion production is possible within stars via nuclear and thermal processes: axions produced in the Sun are called solar axions. These mechanisms and their corresponding couplings are: 

\begin{enumerate}
\item Axion-electron coupling: Atomic, Bremsstrahlung and Compton (ABC)~\cite{Redondo_2013}
\item Axion-nucleon coupling: $^{57}$Fe de-excitation~\cite{Moriyama_1999}
\item Axion-photon coupling: Primakoff effect~\cite{Axion_Book}
\end{enumerate}

\begin{figure}[htbp]
\centering
\includegraphics[width=0.96\linewidth]{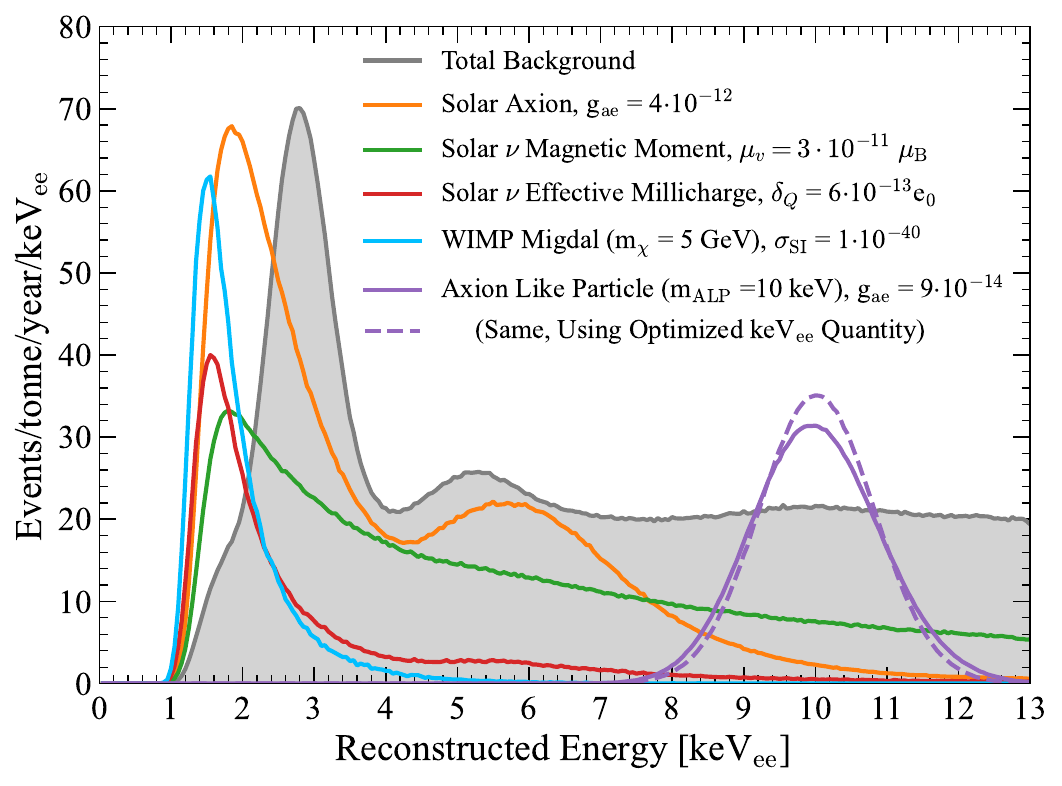}
\caption{Reconstructed energy (keV$_{\mathrm{ee}}$) spectra for the background-only model (grey) and representative signal models: solar axion (orange), solar neutrino magnetic moment (green), solar neutrino millicharge (red), spin independent WIMPs undergoing Migdal effect (blue), and axion like particles (ALPs) (purple).  To produce these spectra, the detector S1 and S2 response is simulated and then all data selections, thresholds, and data quality acceptances are applied.  This standard reconstructed energy quantity can be improved by applying an unequal weighting to the sum of S1$c$ and S2$c$ (see text for details), and we illustrate the advantage of an unequal weighting by showing an alternative reconstruction of the example ALP signal.}
\label{fig:EnergySpectra}
\end{figure}

The interaction of interest in LZ is the axion-electron coupling; this occurs via the axio-electric effect, analogous to the photo-electric effect, allowing atomic ionization by absorption of axions. The axion-electron coupling has a coupling constant given as $g_\mathrm{ae}$.  Following Refs.~\cite{Arisaka_2012,CUORE_axion}, the cross-section for the axio-electric effect is given as
\begin{equation}
     \sigma_{A}=\sigma_{\mathrm{PE}}(E_{A})\frac{{g_{ae}}^{2}}{\beta_{A}}\frac{3{E_{A}}^{2}}{16\pi \alpha {m_{e}}^{2}}(1-\frac{\beta_{A}^{2/3} }{3}),
\end{equation}
where $\sigma_{\mathrm{PE}}$ is the cross-section for the xenon photo-electric effect in barns, $E_{A}$ is the axion total energy, and $\beta_{A}$ is the ratio of the axion velocity to the speed of light. Solar production via the ABC process is dominant compared to the Primakoff effect and $^{57}$Fe de-excitation, and so conservatively only the solar axion flux from the ABC process is considered in this work.  The signature of solar axions in LZ would be an excess of events as seen in Fig.~\ref{fig:EnergySpectra}: an approximately exponentially falling yield coupled to the detector threshold and an L-shell absorption edge of the axio-electric effect leading to a characteristically double peaked spectrum. We limit our searches to solar axion masses $<$1~keV/c$^{2}$ such that their rest mass does not significantly affect the results.

\subsection{Axion-like particles}

The symmetry-breaking framework invoked to solve the strong CP-problem, which motivates the QCD axion, is also used in other BSM theories in which there is breaking of a global U(1) symmetry, with the prediction of pseudo-scalar Nambu Goldstone bosons known as `axion-like' particles (ALPs). Similar light scalars are also predicted to result from higher dimensional gauge fields that are generic in string theories~\cite{alp1, alp2, alp3, alp4, alp5}. In general, ALPs are far less constrained than standard QCD axions, allowing masses and couplings over a much larger parameter space.  As with solar axions, ALPs could be detected in LZ as absorption via the axio-electric effect.  In the case of galactic ALPs, where the velocity is non-relativistic, the spectral signature is a mono-energetic peak (a line feature) corresponding to the ALP rest mass.  An example ALP signal at m$_\mathrm{ALP}$=10keV is shown in Fig.~\ref{fig:EnergySpectra} to illustrate the effect of finite energy resolution.

Following Refs.~\cite{Pospelov_2008,Arisaka_2012} and using factors from Ref.~\cite{Bloch_2016}, if ALPs constitute all of the cold DM in the galaxy with $\rho = 0.3$ GeV$/\mathrm{cm}^{3}$~\cite{DMconventions_2021}, the expected ALP event rate (kg$^{-1}$ day$^{-1}$) in an earthbound detector is given by

\begin{equation}
\label{eqn:ALPrate}
    R_{\mathrm{ALP}} \simeq \frac{1.5\times10^{19}}{A} g^{2}_{\mathrm{ae}}\sigma_{\mathrm{PE}}m_{\mathrm{ALP}},
\end{equation}

\noindent where $A$ is the average mass number of xenon (131.29), $m_{\mathrm{ALP}}$ is ALP mass in keV$/c^{2}$, and $g_{\mathrm{ae}}$ is the ALP-electron coupling constant.

\subsection{Hidden Photons}

The hidden photon (HP) (also called a `dark photon') is a hypothetical U(1)$^\prime$ vector gauge boson within a hidden sector. Hidden photons can obtain a mass through Hidden Higgs or a St{\"u}ckelberg mechanism and interact with the visible sector through loop-induced kinetic mixing with Standard Model hypercharge U(1)$_{\gamma}$ gauge bosons~\cite{Abel_2008}. If HPs were non-thermally produced via the misalignment mechanism, they can reproduce the present-day dark matter relic abundance~\cite{Arias_2012}.

As is the case for ALPs, the absorption of a HP to a bound electron is again analogous to the photo-electric effect, with the photon energy replaced by the HP rest mass {\it m}$_{\mathrm{HP}}$. Following the prescription in Ref.~\cite{Pospelov_2008} and pre-factors in Ref.~\cite{Bloch_2016}, if HP constitutes the DM in the universe, the HP absorption rate (kg$^{-1}$ day$^{-1}$) in a terrestrial detector is given by
\begin{equation}
\label{eqn:HPrate}
    R_{\mathrm{HP}} \simeq \frac{4.7 \cdot 10^{23}}{A} \kappa^{2} \frac{\sigma_{\mathrm{PE}}}{m_{\mathrm{HP}}},
\end{equation}
where $m_{\mathrm{HP}}$ is HP mass in keV$/c^{2}$, and $\kappa$ is the HP kinetic mixing parameter.

\subsection{Midgal Signals from WIMP Recoils}

The Migdal effect refers to the suppressed but nonzero probability of a transition to an ionized atomic state during the initial dark matter scattering process~\cite{MI2018a}.  Through this inelastic scattering process, a NR can appear as a signal that is predominantly of ER character, with a larger fraction of the recoil energy appearing in the observable \{S1$c$, log$_{10}$S2$c$\} signals.  The search for WIMP-induced signals produced via the Migdal effect is now a common experimental method for extending sensitivity to lighter WIMP masses~\cite{LUX_Migdal_2019, EDELWEISS_Migdalsurf_2019, XENON_Migdal_2019, LUX_Migdal_2021, EDELWEISS_Migdal_2022}. It has also been proposed that in response to a recoiling nucleus, a photon could be emitted via Bremsstrahlung. This effect has been quantified in Ref.~\cite{Kouvaris_WIMP_Brem} and considered by LUX in Ref.~\cite{LUX_Migdal_2019}. Bremsstrahlung signals are not considered in this work due to their negligible rate compared to the Migdal process.

As in previous Migdal analyses of nuclear recoil dark matter scattering in Xe~\cite{LUX_Migdal_2019, XENON_Migdal_2019, LUX_Migdal_2021}, the signal model in this analysis relies on Ref.~\cite{MI2018a}. Also in common with previous analyses, we do not include the $n=5$ contributions of the outermost electron shell. We also exclude $n=1$ and $n=2$ contributions because these inner orbitals contribute negligibly to the total spectrum. Alternative calculations of the Migdal effect in Xe typically show good agreement with this signal model, particularly for the $n=3$ and $n=4$ states of interest~\cite{CL2020a, PC2022a}.

When simulating the expected response of LZ to the Migdal signal model, we consider only its ER component, eliminating any discrimination power from the small additional NR component of the recoil. As can be seen in Fig.~\ref{fig:EnergySpectra}, the Migdal signal models are some of the steepest-falling spectra included in this analysis, meaning these models are the most strongly sensitive to models of detector response at threshold.  We discuss this point further in Appendix A.

We also note that dedicated neutron scattering studies are ongoing with the goal of calibrating the Migdal effect in LXe (for $n=2$ and $n=3$)~\cite{migdal_araujo, migdal_bang, migdal_xu}.  Because these calibration efforts are preliminary and ongoing, we note them but do not yet include them in the current work.

\section{Statistical Analysis}
\label{sec:statAnalysis}

In this work, frequentist hypothesis tests based on the profile likelihood ratio (PLR) method are used to exclude signal models described in Section~\ref{sec:signalModels}. The statistical inference of the various model parameters of interest is performed with an unbinned profile likelihood statistic with a two-sided construction of the 90\% confidence bounds as detailed in Refs.~\cite{Cowan_2011, DMconventions_2021}. This work also relies on the same simulation framework as Refs.~\cite{LZ_WIMPresults,LZ_backgrounds}, using the \textsc{geant}4-based package \textsc{baccarat}~\cite{LZ_sims,geant4_2016} and a custom simulation of the LZ detector response using the tuned \textsc{nest} model~\cite{NEST_2011}. For this work, \textsc{nest} version 2.3.12~\cite{NEST_2312} was used.  (While \textsc{nest} version 2.3.7 was used in Refs.~\cite{LZ_WIMPresults,LZ_backgrounds}, comparison tests were performed to ensure v2.3.7 and v2.3.12 give indistinguishable model outputs.)  Detector parameters and model inputs remain unchanged from Refs.~\cite{LZ_WIMPresults,LZ_backgrounds}.  Sensitivity to \textsc{nest} parameter variation is explored in Appendix A.   Background model components receive the same normalizations and constraints as used in Refs.~\cite{LZ_WIMPresults,LZ_backgrounds}.

This work builds on the analysis and statistical framework of LZ's first results~\cite{LZ_WIMPresults}.  The one change to this framework is the inclusion of time-dependence in the likelihood construction:   the included observables are \textbraceleft S1$c$, log$_{10}$S2$c$, time\textbraceright. Given the significant rate of ER backgrounds in the ER signal region of \textbraceleft S1$c$, log$_{10}$S2$c$\textbraceright, the inclusion of time potentially allows for increased sensitivity. Specifically, the rates of the $^{37}$Ar and $^{127}$Xe components are modeled as exponentially falling with their 35.0~d and 36.3~d half-lives, respectively~\cite{TabRad_v8,LUX_Xe127,LZ_Ar37}. The $^{37}$Ar and $^{127}$Xe initial rates are consistent with expectation from above-ground cosmogenic activation (see Refs.~\cite{LZ_Ar37,LZ_backgrounds}) and the subsequent rates are consistent with exponential decay at the expected half-lives (see Ref.~\cite{LZ_backgrounds}).

Time dependence of the $^{222}$Rn daughter $^{214}$Pb was also considered, exploring whether the $^{214}$Pb rate varied significantly over the course of the exposure.  Such variation could potentially arise from, for example, otherwise unnoticed small variation in LXe convection and thermodynamic conditions. The rates of $^{218}$Po and $^{214}$Po (species which precede and follow $^{214}$Pb in the decay chain) were monitored from their alpha decays, and no significant time-dependence was observed.  

Several of the signal models discussed in Section~\ref{sec:signalModels} are expected to exhibit a small time-dependence (for example a $\sim$6\% modulation in the case of solar fluxes of neutrinos and axions). This signal model time-dependence was not included due to the small scales of variation expected and the short ($\sim$5 month) exposure duration. The rate of each signal model was taken to be constant, at the yearly average rate.

The signal or background component PDF is constructed as the product of the appropriate time-independent PDF in \textbraceleft S1$c$, log$_{10}$S2$c$\textbraceright~and the component's time-dependent expected number of counts, g(T).  This time-dependence is itself the product of two terms: 
\begin{equation}
\label{eqn:TimePDF}
\mathrm{g(T)} = \mathrm{t_{live}(T)} \times \mathrm{R(T)},
\end{equation}
where R(T) is the physical rate of the component (either a constant rate or an exponential decay in the case of $^{37}$Ar and $^{127}$Xe), and t$_{\mathrm{live}}$(T) is a histogram of livetime over the course of the exposure binned in 10 minute intervals (much smaller than the timescale of R(T) variation).  The livetime histogram t$_{\mathrm{live}}$(T) integrates to the total livetime. Each background component's g(T) integrates to the total expected background counts as given in Refs.~\cite{LZ_backgrounds,LZ_WIMPresults}.

\begin{figure}[tbp]
  \centering
  \includegraphics[width=0.96\linewidth]{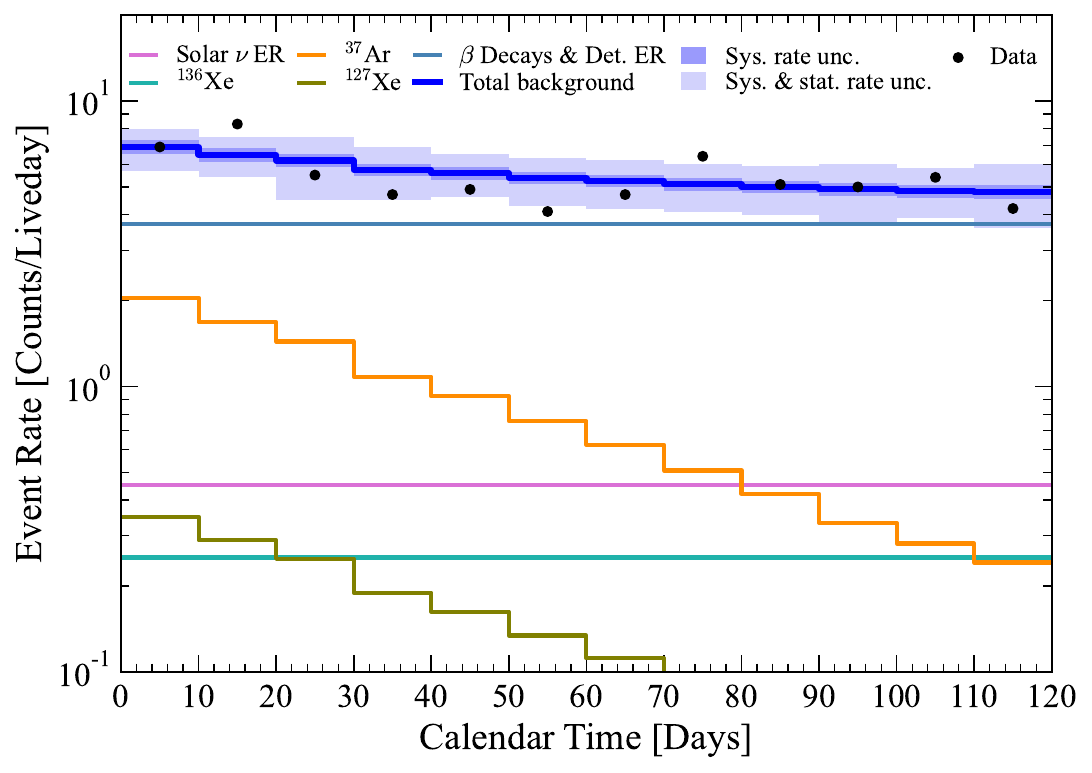}
  \caption{Time dependence of the data and the best-fit model.  Statistical tests are performed on observed and expected counts, while in this plot counts are converted to rates by normalizing by the livetime fraction of each 10-day bin. Data is shown in black. The blue line shows total summed background. The darker blue band shows the 1$\sigma$ model uncertainty and the lighter blue band the combined model and statistical uncertainty. Background components are shown in colors as given in the legend. All background components are included in the fit, but those appearing exclusively below the y-axis bound are not listed in the legend.}
  \label{fig:TimeFit}
\end{figure}

A background-only model fit to data is performed in the observable space \textbraceleft S1$c$, log$_{10}$S2$c$, time\textbraceright.  Figure~\ref{fig:TimeFit} shows this best-fit result projected onto the time axis.  The fitted number of counts for each background component in this 3D space is found to be consistent with the results of fitting in the simpler 2D space \textbraceleft S1$c$, log$_{10}$S2$c$\textbraceright~of Ref.~\cite{LZ_WIMPresults}, and the goodness-of-fit when projected onto the keVee axis is similarly retained.  When the background-only fit is projected onto the time axis as in Fig.~\ref{fig:TimeFit}, the p-value is 0.43.  When the ROI is restricted to S1$c$ \textless~20, the p-value is 0.35.  A WIMP search was performed using this new time-dependent framework, to check consistency with the time-independent result of Ref.\cite{LZ_WIMPresults}.  As expected, the improvement in limit was small ($<$5\%) due to the small number of background counts expected in the NR signal regions in \textbraceleft S1$c$, log$_{10}$S2$c$\textbraceright.

\section{Search Results}
\label{sec:Results}

In this section we report the results of searches for the signal models discussed in Section~\ref{sec:signalModels}. For all signal models considered, the p-value of the background-only hypothesis is greater than 10\%. Figures~5-9 show 90\% confidence level (C.L.) upper limits.  In each figure, the solid black line is the observed upper limit, and the dashed black line is the median sensitivity. The green and yellow bands represent the 1$\sigma$ and 2$\sigma$ sensitivity range, respectively, for repeated background-only experiments. 

\begin{figure}[!tbp]
    \centering
    \includegraphics[width=0.99\linewidth]{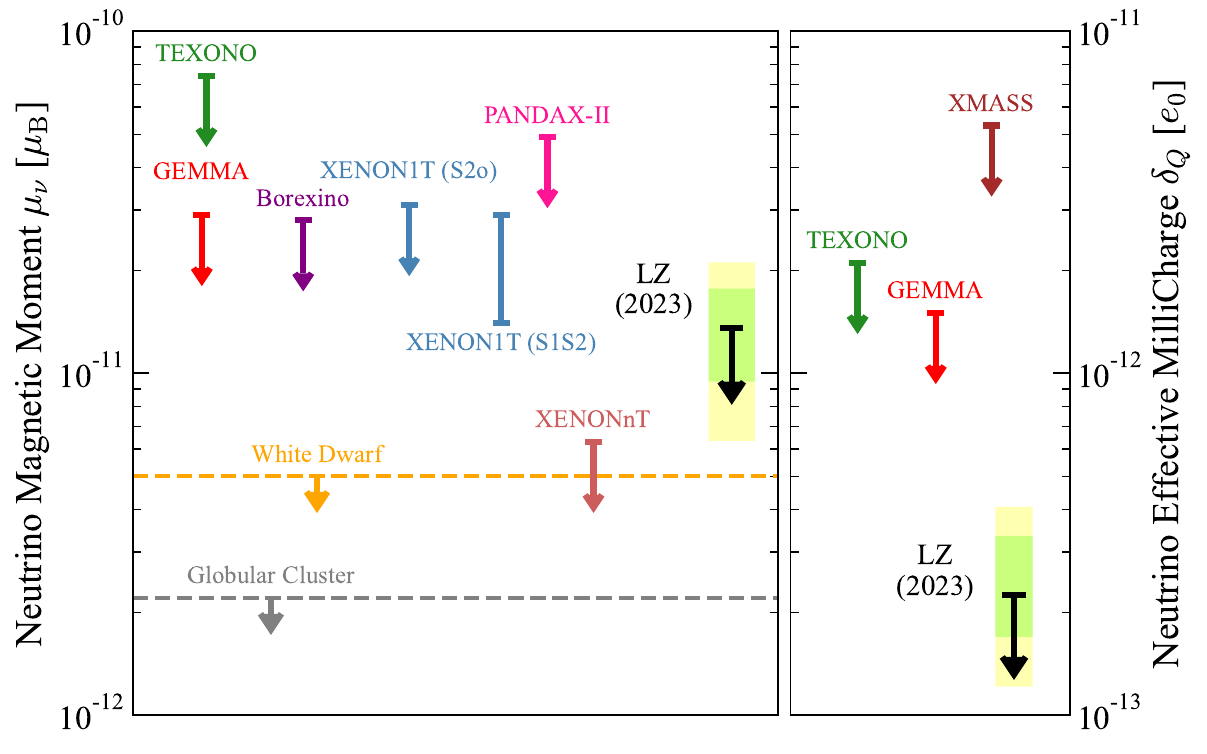}
    \caption{The 90\% C.L. upper limit on effective neutrino magnetic moment (left) and neutrino effective millicharge (right). Selected limits from other experiments are also shown~\cite{XENON1T_excess, TEXONO_NuMM_2009,GEMMA_NuMM_2009,Borexino_NuMM_2017,PandaXII_NuMM_2020,XMASS_NuMM_2005,XENON_ERlimits_2022} and astrophysical observations  \cite{Corsico_whiteDwarf,Viaux_globclusters}.}
    \label{fig:limits_SolarNu}
\end{figure}

Figure~\ref{fig:limits_SolarNu} shows search results for the neutrino effective magnetic moment and effective millicharge, parameterized in terms of the Bohr magneton, $\mu_{B}$, and electric charge, $e_{0}$, respectively. The observed 90\% C.L. upper limit for the neutrino effective magnetic moment is 1.36 $\times$ 10$^{-11}$ $\mu_{B}$. Currently, the most stringent constraints on the neutrino magnetic moment are from astrophysical measurements of cooling rates of pulsating white dwarfs \cite{Corsico_whiteDwarf} and from high precision photometry of red-giant branch stars in globular clusters~\cite{Viaux_globclusters}. The observed 90\% C.L. upper limit for the neutrino effective millicharge is 2.24 $\times$ 10$^{-13}$ $e_{0}$. We note that, using the released data-set and \textsc{nest} detector parameters from LZ's first science run in Ref.~\cite{LZ_WIMPresults}, the authors in Ref.~\cite{Corona_SolarNu_2022} independently set 90\% C.L. limits on effective neutrino magnetic moment of  1.1 $\times$ 10$^{-11}$ $\mu_{B}$ and effective neutrino millicharge of 1.5 $\times$ 10$^{-13}$ $e_{0}$.  These limits of Ref.~\cite{Corona_SolarNu_2022} are strengthened by considering only a single combined uncertainty on all background rates (rather than component-by-component), and further by underestimating that combined uncertainty.

\begin{figure}[!tbp]
    \centering
    \includegraphics[width=0.96\linewidth]{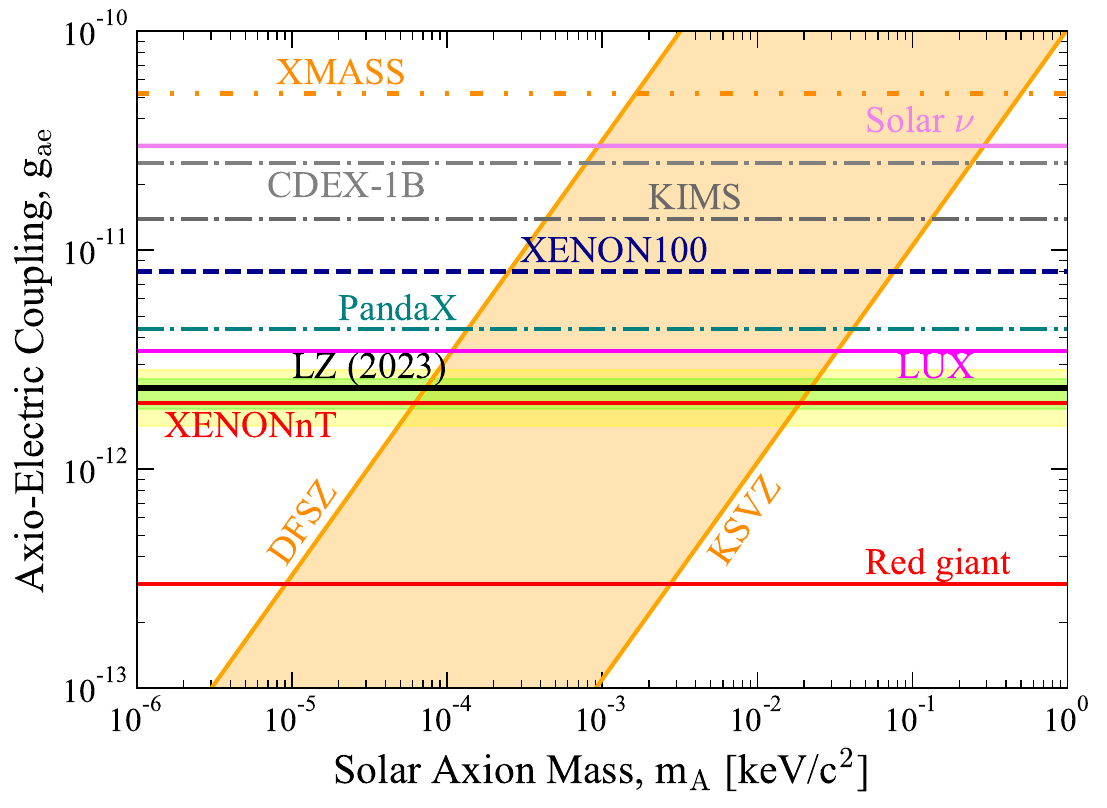}
    \caption{The 90\% C.L. upper limit (black line) on solar axion g$_{\mathrm{ae}}$ coupling constant. Selected limits from other experiments and astrophysical observations are also shown \cite{XENON_ERlimits_2022, RedGiant_gae, XMASS_gae, PandaX_axion, solar_neutrino_gae, CDEX_gae, KIMS_axion, LUX_gae}. The shaded orange region corresponds to predicted values from the benchmark QCD axion models DFSZ~\cite{Dine_1981,Zhitnitsky_1980} and KSVZ~\cite{Kim_1979,Shifman_1979}. The LZ median sensitivity is not displayed due to close overlap with the observed limit.}
    \label{fig:SolarAxion_limits}
\end{figure}

Figure~\ref{fig:SolarAxion_limits} shows search results for solar axion coupling, $g_{\mathrm{ae}}$, valid across the mass range shown. The observed 90\% C.L. upper limit for the solar axion coupling constant is $g_{\mathrm{ae}} = 2.35\times10^{-12}$. The current strongest constraint on $g_{\mathrm{ae}}$ is from astrophysical constraints extrapolated from cooling rates of red giants at a level of $g_{\mathrm{ae}} = 3\times10^{-13}$~\cite{RedGiant_gae}. In this first LZ dataset, the solar axion spectrum would overlap with background contributions from the 2.82~keV peak of $^{37}$Ar and the 5.2~keV peak of $^{127}$Xe.

\begin{figure*}[!th]
    \noindent
    \centering
    \subfloat{\includegraphics[width=.45\textwidth]{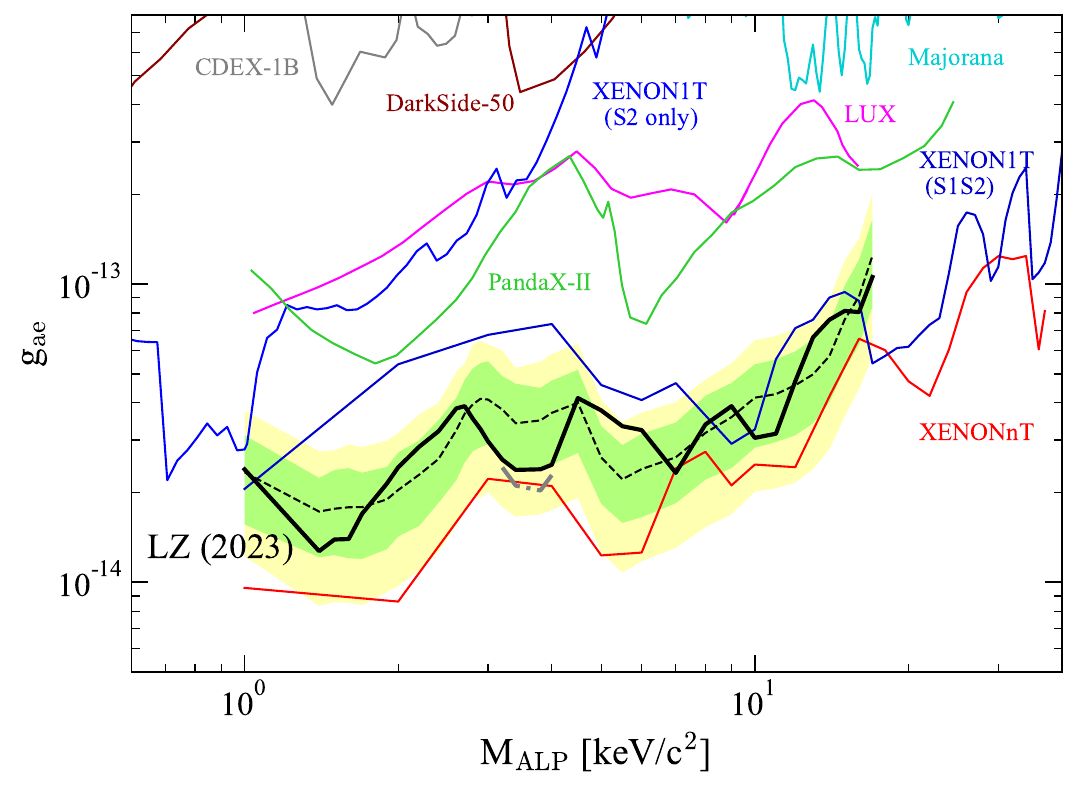}}
    \hfill
    \centering
    \subfloat{\includegraphics[width=.45\textwidth]{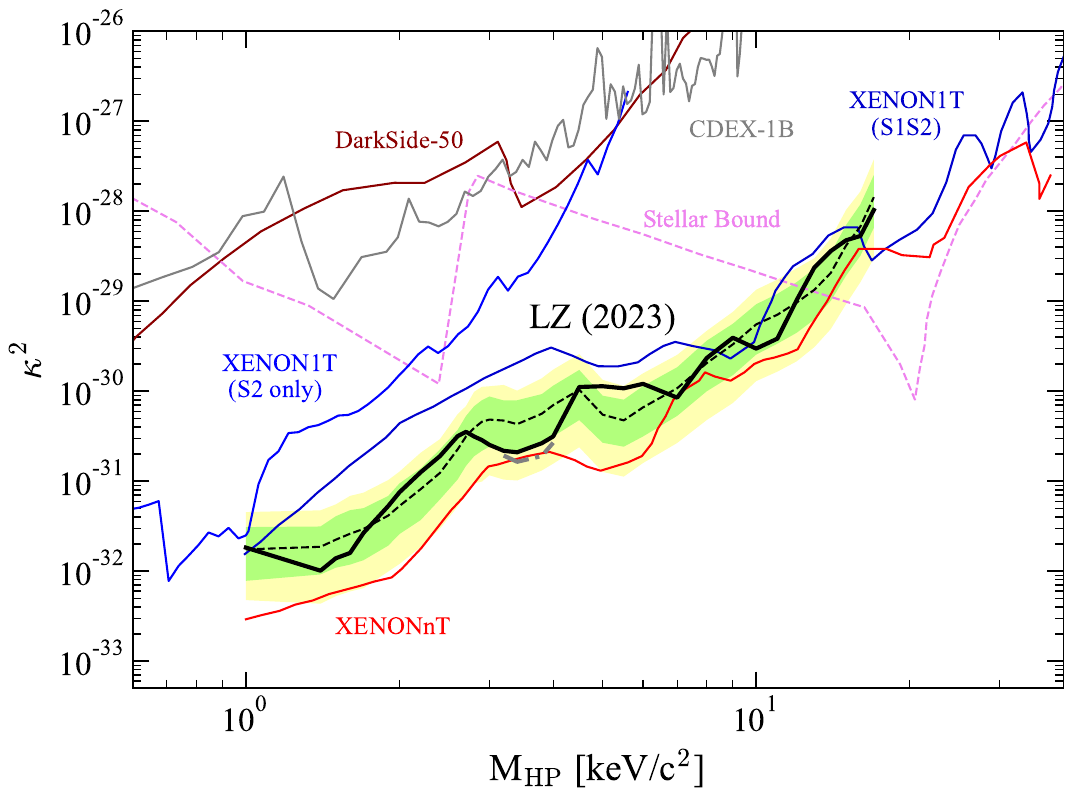}}
    \caption{The 90\% C.L. upper limit (black line) on ALP coupling constant, $g_{\mathrm{ae}}$ (left panel), and the HP coupling constant squared, $\kappa^{2}$ (right panel). Selected limits from other experiments are also shown \cite{XENON_S2o_limits, XENON1T_excess, XENON_ERlimits_2022, CDEX_gae, GERDA_ALPs, LUX_gae, Majorana_ALPs, PandaX_axion, darksideER} along with astrophysical bounds in Ref.~\cite{Haipeng_HP_stellarBounds}.}
    \label{fig:MonoE_limits}
\end{figure*}

Figure~\ref{fig:MonoE_limits} shows search results for mono-energetic signals of ALPs and HPs, as a function of mass across the low energy ROI. ALP and HP masses spanning from 1~keV to 17~keV are considered. Sensitivity near the 2.82~keV energy of $^{37}$Ar is increased through the inclusion of time-dependence. Between 3.2 keV and 4 keV, random fluctuation in the data results in upper limits which fall below the median sensitivity by more than 1$\sigma$. For this specific mass interval, we employ the Power Constrained Limit method~\cite{Cowan_2011b} to not report a limit stronger than the $-1\sigma$ range of the projected sensitivity.  We note that the XENONnT searches of Ref.~\cite{XENON_ERlimits_2022} do not apply a similar constraint.

\begin{figure*}
\begin{minipage}[!t]{2\columnwidth}
  \centering
  \includegraphics[width=0.5\textwidth]{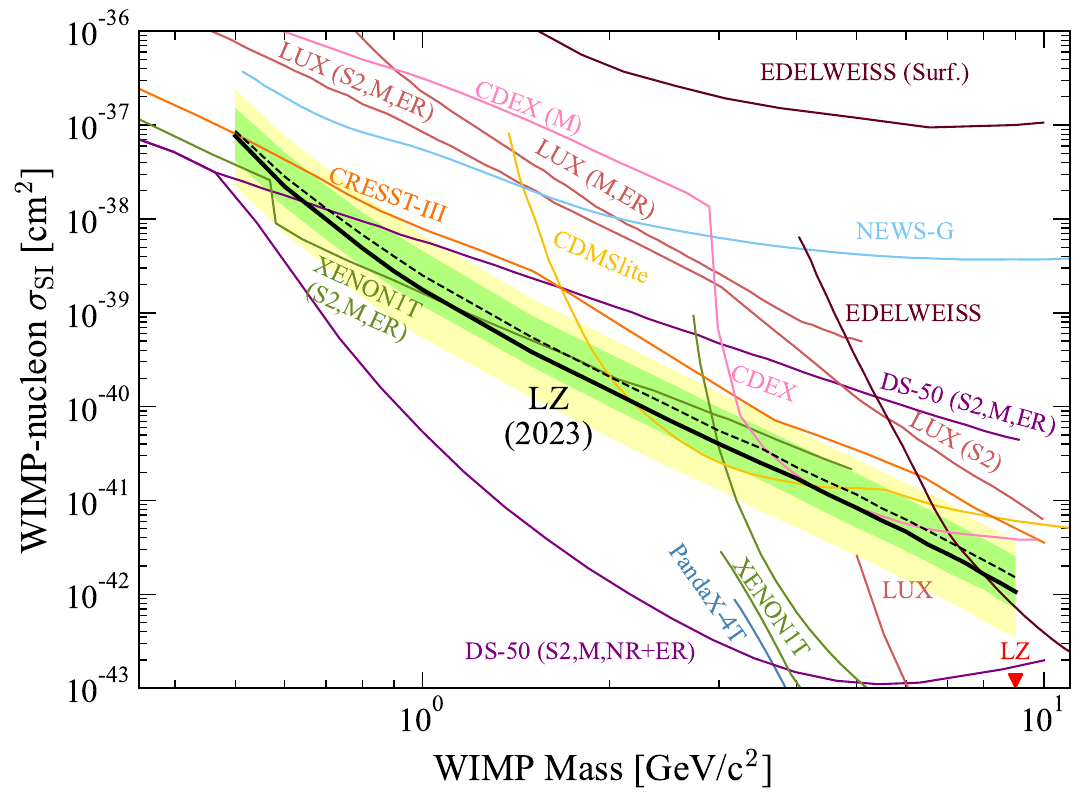}
  \caption{The 90\% C.L. upper limit (black line) for the spin-independent (SI) WIMP cross-section vs. WIMP mass, using the Migdal signal pathway. Also shown are limits from other experiments \cite{LUX_Migdal_2019,LUX_Migdal_2021,EDELWEISS_Migdalsurf_2019,XENON_S2o_limits,XENON_Migdal_2019,EDELWEISS_Migdal_2022,CDEX_Migdal_2019,CRESST_Migdal_2019,CDMSlite_Migdal_2017,NEWSG_DM_2018,CDMS_S2only_2016,DarkSide_S2only_2018,DarkSide_Migdal_2022,DarkSide-50:2022qzh, PandaX-4T_SI-DM_2021,SuperCDMS_CPD_2021,LUX_2016DMresults,XENON_SI-DM_2018}.  Labeling indicates limits employing a threshold in the S2-only regime (`S2'), a Migdal recoil process (`M'), and/or a conservative signal Migdal model including only the ER component (`ER').  The DarkSide-50 ER+NR Migdal result employs the `QF' liquid Ar NR response model described in~\cite{DarkSide-50:2022qzh,darkside_calibration}, which assumes a Lindhard electronic partition, binomial quenching fluctuations, and no recombination enhancement due to the proximity of the ER and NR components.}
  \label{fig:MigdalSI_limits}
\end{minipage}
\end{figure*}

\begin{figure*}[!ht]
    \centering
    \subfloat{\includegraphics[width=.45\textwidth]{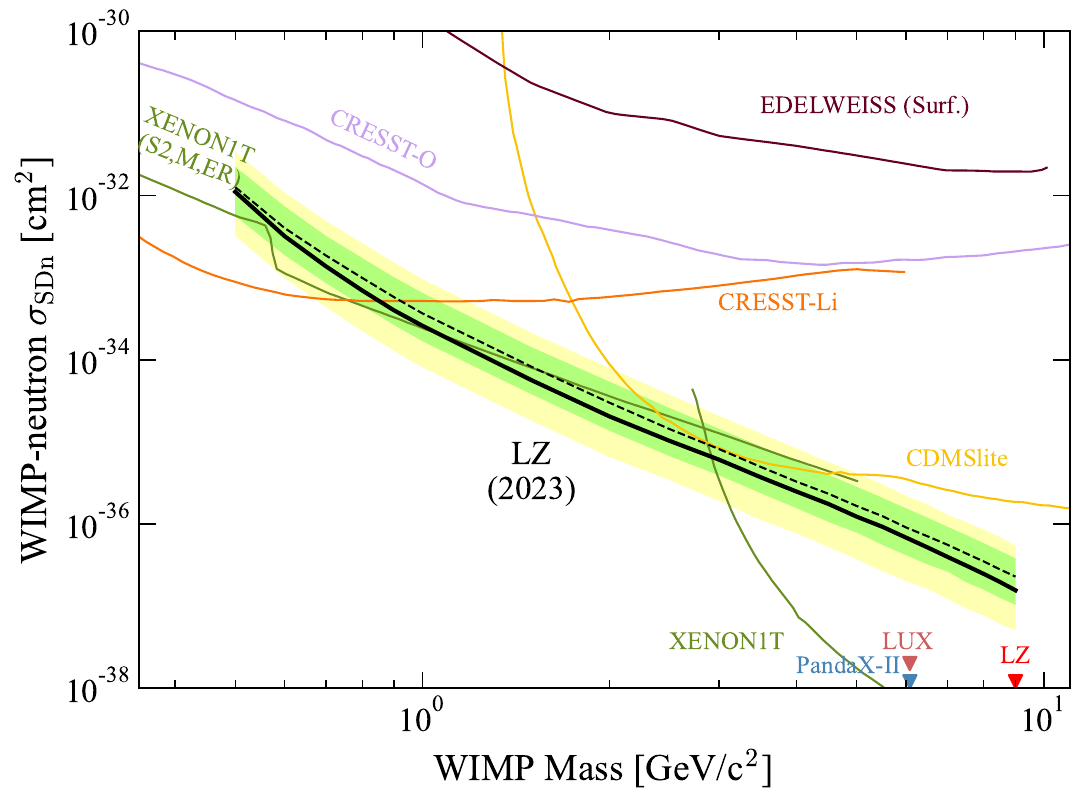}}
    \hfill
    \subfloat{\includegraphics[width=.45\textwidth]{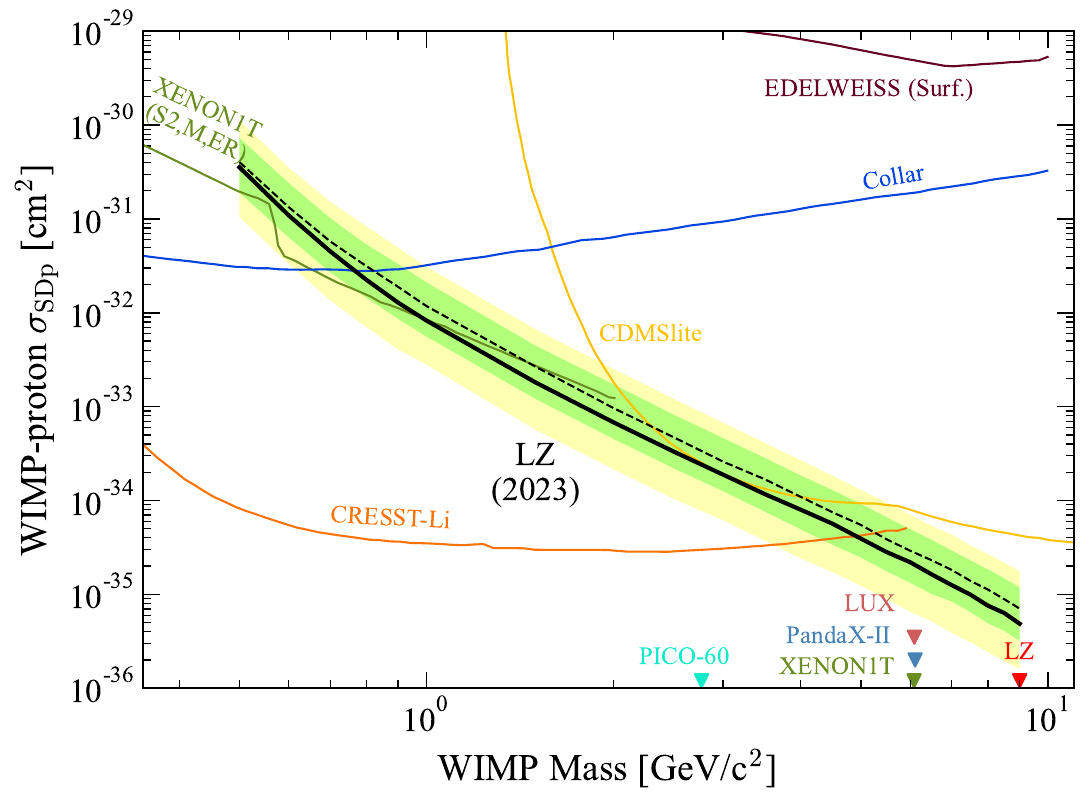}}
    \caption{The 90\% C.L. upper limit (black line) for the spin-dependent (SD) WIMP cross-section versus WIMP mass for coupling on neutrons (left) and protons (right) undergoing the Migdal effect. The SD proton and neutron modes use the mean nuclear structure functions from \cite{Hoferichter_2020}. Also shown are limits from other experiments \cite{XENON_S2o_limits,CDMSlite_Migdal_2017,Collar_SD_2018,EDELWEISS_Migdalsurf_2019,XENON_Migdal_2019,CRESST_Migdal_2019,CRESST_Li_SD_2019,CRESST_li_SD_2022}.}
    \label{fig:MigdalSD_limits}
\end{figure*}

Figures~\ref{fig:MigdalSI_limits} and~\ref{fig:MigdalSD_limits} show search results for WIMPs undergoing the Migdal effect from 0.5 to 9 GeV/c$^{2}$. Masses below 0.5 GeV/c$^{2}$ were not considered in this work as the energies are predominantly sub-threshold and rarely fluctuate above threshold so as to produce S1 signals which are both $\ge$3-fold and $>$3~phd. 90\% C.L. upper limits are shown for both spin-independent (SI) and spin-dependent (SD) coupling modes. For the SD-proton and SD-neutron modes, the mean nuclear structure functions from Ref.~\cite{Hoferichter_2020} were used to ensure consistency with limits from previous xenon-based limits. Based on these results, LZ has placed strong limits upon \textless~2~GeV/c$^{2}$ WIMPs undergoing a SI coupling mode, and \textless~3~GeV/c$^{2}$ WIMPs undergoing a SD-neutron coupling mode. The SD-neutron coupling is dominated by the unpaired neutrons of $^{129}$Xe (spin 1/2, 26.4\% abundance) and $^{131}$Xe (spin 3/2, 21.2\% abundance)~\cite{Xe_abundances}.

We note that SD WIMP interactions have significant uncertainties due to nuclear modeling, and these uncertainties are not included in the quoted results or shown in Fig.~\ref{fig:MigdalSD_limits}. These nuclear uncertainties can be estimated (for each WIMP mass) by calculating the global minimum and maximum interaction rate at each energy across Refs.~\cite{Hoferichter_2020, Hu_2021, Pirinen_2019}.  For the SD-neutron (SD-proton) coupling in Xe, the global minimum and maximum nuclear coupling models lead to limits changing by, on average, factors of 0.74 to 2.7 (0.81 to 44).

As described in Secs. II and IV, the present work has included the time-dependence of the $^{37}$Ar and $^{127}$Xe background contributions. To measure the impact on observed limits of this inclusion of time-dependence, the same statistical analysis was performed with all time-dependence removed. As expected, the largest impact is on monoenergetic signals near the $^{37}$Ar 2.82~keV (K-shell) peak.  For example, for 2.8 keV ALPs, the observed limit is weakened by a factor of 3.  The inclusion of time-dependence enhances sensitivity by a smaller amount for the other signals considered, in proportion to their overlap with the $^{37}$Ar background (the dominating time-dependent background):  11.5\% for neutrino magnetic moment, 5.1\% for neutrino effective millicharge, 9.2\% enhancement for solar axions, and \textless10\% across WIMP masses undergoing Migdal effect.

\section{Conclusions}

In summary, we have performed eight BSM physics searches in low-energy electron recoil signals using the same 0.91 tonne$\times$years exposure and data selections as presented in the experiment's first WIMP results~\cite{LZ_WIMPresults}.  In the case of SD-neutron Migdal and neutrino effective millicharge, the upper limits exclude new regions of their respective parameter space.  Where this work tests models already tested by XENONnT~\cite{XENON_ERlimits_2022}, a difference in $^{214}$Pb rate gives XENONnT comparatively greater sensitivity for a similar exposure.

LZ is continuing to collect data after a period of calibrations and detector state optimizations. Subsequent science data will benefit from further decay of both $^{127}$Xe and $^{37}$Ar, which will especially improve the solar axion sensitivity.  Separately, it has been demonstrated that adjustments to LXe flow in LZ can reduce the rate of $^{214}$Pb in the fiducial volume, improving sensitivity to all ER signals.  Because several signal spectra are steeply falling near threshold, reducing threshold in analysis will provide increased sensitivity, especially to WIMPs undergoing Migdal effect and to neutrino electromagnetic moments.  Such threshold reductions can be enacted by accepting smaller S1 amplitudes or by switching to an S2-only search.  Through continued data-taking with reduced backgrounds, and analysis methods with lower thresholds, we expect significant future advancement in sensitivities to ER signals.

\begin{acknowledgments}

The research supporting this work took place in part at SURF in Lead, South Dakota. Funding for this work is supported by the U.S. Department of Energy, Office of Science, Office of High Energy Physics under Contract Numbers DE-AC02-05CH11231, DE-SC0020216, DE-SC0012704, DE-SC0010010, DE-AC02-07CH11359, DE-SC0012161, DE-SC0015910, DE-SC0014223, DE-SC0010813, DE-SC0009999, DE-NA0003180, DE-SC0011702, DE-SC0010072, DE-SC0015708, DE-SC0006605, DE-SC0008475, DE-SC0019193, DE-FG02-10ER46709, UW PRJ82AJ, DE-SC0013542, DE-AC02-76SF00515, DE-SC0018982, DE-SC0019066, DE-SC0015535, DE-SC0019319, DE-AC52-07NA27344, \& DOE-SC0012447.	This research was also supported by U.S. National Science Foundation (NSF); the UKRI’s Science \& Technology Facilities Council under award numbers ST/M003744/1, ST/M003655/1, ST/M003639/1, ST/M003604/1, ST/M003779/1, ST/M003469/1, ST/M003981/1, ST/N000250/1, ST/N000269/1, ST/N000242/1, ST/N000331/1, ST/N000447/1, ST/N000277/1, ST/N000285/1, ST/S000801/1, ST/S000828/1, ST/S000739/1, ST/S000879/1, ST/S000933/1, ST/S000844/1, ST/S000747/1, ST/S000666/1, ST/R003181/1; Portuguese Foundation for Science and Technology (FCT) under award numbers PTDC/FIS-PAR/2831/2020; the Institute for Basic Science, Korea (budget number IBS-R016-D1). We acknowledge additional support from the STFC Boulby Underground Laboratory in the U.K., the GridPP~\cite{faulkner2005gridpp,britton2009gridpp}  and IRIS Collaborations, in particular at Imperial College London and additional support by the University College London (UCL) Cosmoparticle Initiative. We acknowledge additional support from the Center for the Fundamental Physics of the Universe, Brown University. K.T. Lesko acknowledges the support of Brasenose College and Oxford University. The LZ Collaboration acknowledges key contributions of Dr. Sidney Cahn, Yale University, in the production of calibration sources. This research used resources of the National Energy Research Scientific Computing Center, a DOE Office of Science User Facility supported by the Office of Science of the U.S. Department of Energy under Contract No. DE-AC02-05CH11231. We gratefully acknowledge support from GitLab through its GitLab for Education Program. The University of Edinburgh is a charitable body, registered in Scotland, with the registration number SC005336. The assistance of SURF and its personnel in providing physical access and general logistical and technical support is acknowledged. We acknowledge the South Dakota Governor's office, the South Dakota Community Foundation, the South Dakota State University Foundation, and the University of South Dakota Foundation for use of xenon. We also acknowledge the University of Alabama for providing xenon.  For the purpose of open access, the authors have applied a Creative Commons Attribution (CC BY) licence to any Author Accepted Manuscript version arising from this submission.

\end{acknowledgments}

\section*{Appendix A: Signal Yield at Threshold}
\label{appendixA}

Of the signal models tested in this work, several spectra are steeply falling near threshold.  (This threshold is imposed by S1 selection requirements:  at least 3 PMTs must record some signal, and the summed signal must be greater than 3~phd.)  For steeply falling signals, variation of the \textsc{nest} LXe response model can lead to variation in signal detection efficiency, which can then result in noticeable changes on total signal rate. In this appendix we vary the LXe response model at low energies and quantify the possible impact on the limits presented in this work.

As discussed in Sec.~\ref{sec:detector} and Ref.~\cite{LZ_WIMPresults}, the detector and LXe response parameters of the \textsc{nest}~\cite{NEST_237} ER model were fitted to match the median and widths of tritium calibration data in \{S1$c$, log$_{10}$S2$c$\}, and to simultaneously match the reconstructed energies of the $^{83m}$Kr (41.5 keV), $^{129m}$Xe (236 keV), and $^{131m}$Xe (164 keV) peaks. While this response model parameter fitting was performed using \textsc{nest} 2.3.7, checks were performed to show that the \textsc{nest} 2.3.12 used in this work gave indistinguishable outputs.  \textsc{nest} 2.3.12 includes functionality allowing for ER and NR charge yield parameters to be easily varied. The fit results in small uncertainties on the S1$c$ and S2$c$ gain parameters ($g_1$ = 0.114$\pm$0.002 phd/photon and $g_2$ = 47.1$\pm$1.1 phd/electron). Variation of $g_1$ and $g_2$ within these 1$\sigma$ ranges was found to have a negligible effect on sensitivities or observed limits. 

The more relevant uncertainty is the model of initial charge yield, Q$_{y}$, for ER response at low energies.  The Q$_{y}$ model function is a sum of two sigmoids, parameterized as
\begin{align*}
\label{eqn:BetaQy}
    \mathcal{\mathrm{Q}}_{y}(E,\mathcal{E},\rho) =\hspace{3mm} &m_{1}(\mathcal{E},\rho) &+\hspace{3mm} &\frac{m_{2} - m_{1}(\mathcal{E},\rho)}{[1+(\frac{E}{m_{3}})^{m_{4}}]^{m_9}}\\
                                                               &+\hspace{2mm} m_{5}     &+\hspace{3mm} &\frac{- m_{5}}{[1+(\frac{E}{m_{7}(\mathcal{E})})^{m_{8}}]^{m_{10}}},
\end{align*}

\noindent where $E$ is the recoil energy, $\mathcal{E}$ is the drift field, $\rho$ is the LXe density, and the parameters $m_{1}$ through $m_{10}$ are fit to world data~\cite{NEST_reviewModels}.  In this parameterization, Q$_{y}$ asymptotically approaches a fixed value at low energies, and $m_2$ sets that low-energy asymptotic value. The value of $m_2$ is constrained by world data to $m_2=77.3\pm8.0$~keV$^{-1}$ (electrons per keV).  At low energies, charge yield, Q$_{y}$, is more robustly constrained experimentally than the corresponding scintillation light yield, L$_{y}$, and so the Q$_{y}$ function forms the foundation of the \textsc{nest} description. Assuming a fixed overall work function, any increase in charge yield Q$_{y}$ corresponds to an equal decrease in the light yield L$_{y}$, which can be directly calculated.

To further constrain Q$_{y}$ near threshold, Q$_{y}$ was fit to LZ tritium calibration data.  Given that the energy range in question was specifically the charge yield at threshold, only the low-energy ($\lesssim$2.5~keV) portion of Q$_{y}$ was varied in the fit, while the higher-energy ($\gtrsim$2.5~keV) portion remained fixed according to world data.  This was accomplished by varying $m_2$ as the fit parameter, and keeping all other parameters fixed at their globally-constrained values, with one exception:   $m_3$ was varied in a 1-to-1 correlated manner with $m_2$ to keep Q$_{y}$ unvaried for E$\gtrsim$2.5~keV.  The result of this fit procedure was a best-fit value of $m_2$ of $85.0\pm10.0$~keV$^{-1}$ (with corresponding $m_3$ of 0.73$\pm$0.16~keV).

\begin{figure}[htbp]
  \centering
  \includegraphics[width=0.96\linewidth]{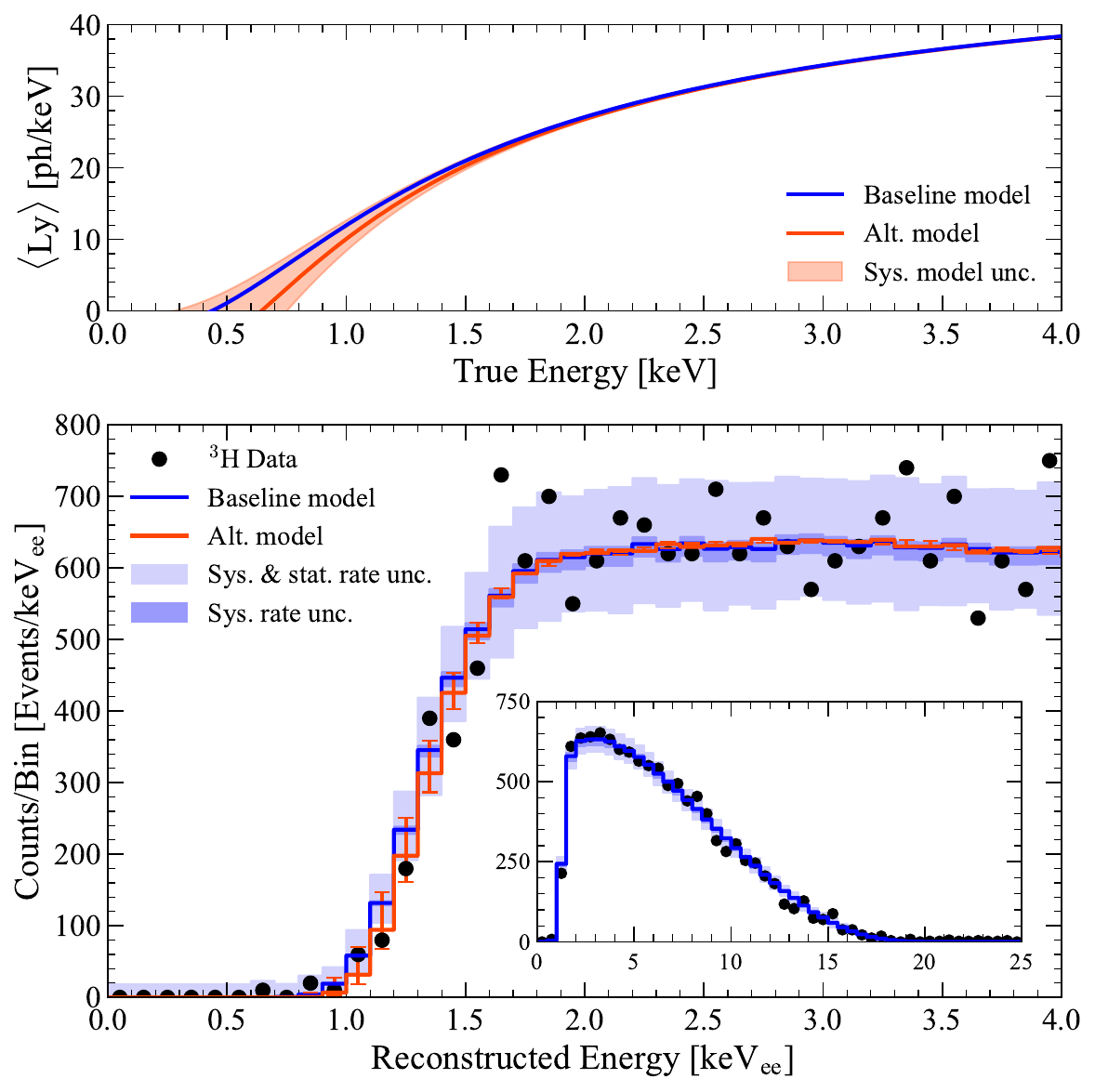}
  \caption{Top panel:   The `baseline' (global fit) LXe response model and the `alternative' (LZ tritium-fitted) response models, as referred to in this section, both illustrated as light yields L$_y$. The fit uncertainty on the alternative model is also shown.  Lower panel:  Reconstructed energy spectra of the tritium calibration source is shown (black) after all cuts are applied, along with the baseline model (blue) as described in \cite{LZ_WIMPresults}. The dark blue band represents systematic uncertainty from data quality acceptances and the light blue band is the Poissonian statistical uncertainty (added to systematic uncertainty).  In red is the alternative response model. The red error bars illustrate the uncertainties in the model fit, corresponding to the shaded red region in the upper panel. Note that the same systematic and statistical uncertainties on the baseline model would also apply to the alternative model.}
  \label{fig:TritumTuning}
\end{figure}

L$_{y}$ models corresponding to the fitted Q$_y$ models are shown in the upper panel of Fig.~\ref{fig:TritumTuning} for both the `baseline' globally-constrained response model used in this work (blue) and the `alternative' model constrained by the LZ tritium data as described above (red).  These two models are both compared with the tritium calibration data in the lower panel.  It can be seen that the alternative response model results in a small improvement in $\chi^{2}$ in the threshold region of interest:   In the 0-2.5 keV$_{\mathrm{ee}}$ region, the baseline model p-value is 0.655, while the alternative model p-value is 0.785. The ER threshold, as shown in Figure.~\ref{fig:efficiency}, is set by the S1 amplitude. The higher estimate for Q$_{y}$ at low energies corresponds to a lower estimate for L$_{y}$, leading to a slightly elevated threshold (a smaller fraction of near-threshold events produce an S1 signal above threshold).  Thus any limits produced using this alternative response model will lead to reduced signal rates and should be considered conservative with respect the baseline model. The best-fit and uncertainty in L$_{y}$ is represented by the red line and red band, respectively, within the top panel of Figure.~\ref{fig:TritumTuning}, and its corresponding impact on ER detection efficiency is shown as the purple band in Figure.~\ref{fig:efficiency}.

With the alternative ER response model in hand, new `alternative' models for the signals and backgrounds in $\{$S1$c$, log$_{10}$S2$c\}$ were constructed, and within this alternative response model framework the identical profile-likelihood procedure was repeated to estimate the effect on sensitivities and observed limits. Two separate effects play a role in any change: 1) the signal rate can change due to loss of efficiency at threshold and 2) the signal shape can change in $\{$S1$c$, log$_{10}$S2$c\}$. 

In the case of solar axions, the observed limit on coupling strength, g$_{\mathrm{ae}}$, is weakened by 3.1\%. In the case of the solar neutrino magnetic moment and solar neutrino millicharge signals, the alternative model weakens the observed limits by 8.7\% and 18.9\%, respectively.  The difference reflects their differing spectral shapes of $\sim$E$_r^{-1}$ and $\sim$E$_r^{-2}$, respectively.

The alternative response model has only a small impact on the observed limits for solar axions and monoenergetic signals (ALPs and HPs) with masses above 2~keV/c$^{2}$, where threshold effects are less important. For ALPs of 1.5 keV/c$^{2}$ mass, the limit on g$_{\mathrm{ae}}$ is weakened by 11.9\%.  For HPs of a similar mass, the limit on $\kappa^{2}$ is weakened by 25.3\%.  ALP and HP searches were truncated at 1.0~keV to keep this response model rate variation $<$50\%. 

The Migdal spectra are also steeply falling at threshold, in particular for the lowest WIMP masses. As in the ALP and HP case, we restrict our search to a mass range for which the alternative response model reduces the signal rate by $<$50\%.  In the Migdal case, we find this is true for all masses above 0.5~GeV/c$^{2}$ (and is similar for all three coupling types: SI, SDn, and SDp). 

In conclusion, sensitivity to ER signals which appear primarily at threshold can be impacted by uncertainty in the low-energy LXe response model (parameterized within \textsc{nest}).  We estimate the scale of this uncertainty and restrict our reported limits to models for which the uncertainty is small.  A similar response uncertainty is common to all S1+S2 LXe searches, and such response model uncertainties should be considered when reporting observations in the S1 threshold region.

\section*{Appendix B: Data Release}
\label{appendixB}

\noindent Selected data from from this paper are available\\ at
\url{https://tinyurl.com/LZDataReleaseRun1ER}

\begin{itemize}
    \item Figure 1:  The search data after all analysis selections applied, in a format of $\{$S1$c$, log$_{10}$S2$c$, time since start of SR1$ \}$ (black dots).
    \item Figure 2:  Points representing the total ER efficiency curve for this analysis (black line).
    \item Points representing a histogram of livetime over the time window of this analysis.
    \item Figures 7, 8, 9:  Points representing the observed 90\% confidence level upper limits, together with the median, $\pm1\sigma$, and $\pm2\sigma$ expected sensitivities.
\end{itemize}

\newpage

\bibliographystyle{apsrev4-2}
\bibliography{reference}

\end{document}